\documentclass[journal,comsoc]{IEEEtran}
\usepackage[T1]{fontenc}

\ifCLASSINFOpdf
\else
\fi
\usepackage{cite}
\usepackage{amsmath,amssymb,amsfonts}
\usepackage{algorithm,algorithmic}
\usepackage{graphicx}
\usepackage{xcolor}
\usepackage{algorithmic}
\usepackage{graphicx}
\usepackage{caption}
\usepackage{amsmath}
\usepackage{algorithmic}
\usepackage{tabu}
\usepackage{lineno}
\usepackage{amsthm}
\usepackage{mathtools}
\usepackage{footnote}

\usepackage{caption}
\usepackage{amsmath}
\usepackage{subcaption}
\usepackage{amsmath, bm}

\DeclareMathAlphabet{\mathcal}{OMS}{cmsy}{m}{n}

\usepackage[cmintegrals]{newtxmath}
\begin{document}
\title{Trustworthy Artificial Intelligence Framework for Proactive Detection and Risk Explanation of Cyber Attacks in Smart Grid}

	\author{Md.~Shirajum~Munir,~\IEEEmembership{Member,~IEEE,}
		Sachin~Shetty, \IEEEmembership{Senior~Member,~IEEE,}
		and~Danda B. Rawat,~\IEEEmembership{Senior~Member,~IEEE}
		\thanks{Md. Shirajum Munir and 	Sachin Shetty are with the Virginia Modeling, Analysis, and Simulation Center, Old Dominion University, Suffolk, VA 23435, USA(e-mail: mmunir@odu.edu; sshetty@odu.edu).}
		\thanks{Danda B. Rawat is with the Department of Electrical and Computer Science,Howard University, 2400 Sixth Street NW, Washington, D.C. 20059, USA (e-mail: rawat@howard.edu).}
		\thanks{This work is supported in part by DoD Center of Excellence in AI and Machine Learning (CoE-AIML) under Contract Number W911NF-20-2-0277 with the U.S. Army Research Laboratory.}
} 

	\markboth{Submitted for peer review. Copyright subject to be reserved to IEEE}%
	{Shell \MakeLowercase{\textit{et al.}}: Bare Demo of IEEEtran.cls for IEEE Communications Society Journals}
	\maketitle
	
\begin{abstract}
The rapid growth of distributed energy resources (DERs), such as renewable energy sources, generators, consumers, and prosumers in the smart grid infrastructure, poses significant cybersecurity and trust challenges to the grid controller. Consequently, it is crucial to identify adversarial tactics and measure the strength of the attacker's DER. To enable a trustworthy smart grid controller, this work investigates a trustworthy artificial intelligence (AI) mechanism for proactive identification and explanation of the cyber risk caused by the control/status message of DERs. Thus, proposing and developing a trustworthy AI framework to facilitate the deployment of any AI algorithms for detecting potential cyber threats and analyzing root causes based on Shapley value interpretation while dynamically quantifying the risk of an attack based on Ward's minimum variance formula. The experiment with a state-of-the-art dataset establishes the proposed framework as a trustworthy AI by fulfilling the capabilities of reliability, fairness, explainability, transparency, reproducibility, and accountability.
\end{abstract}
	

	\IEEEpeerreviewmaketitle

\section{INTRODUCTION}
\label{sec:intro}
The smart grid infrastructure is expected to deploy huge amounts of distributed energy resources (DERs) such as renewable energy sources, consumers, prosumers, and so on to meet the goal of around $40\%$ \cite{EnergyOutlook2023} cost reduction by $2050$. Thus, the operation of such deployed DERs in a smart grid significantly increases the risk of cyber-attacks through their control/status messages \cite{SmartgridAttack4,SmartgridAttack3,SmartgridAttack1,SmartgridAttack2}. As a result, it is critical to recognize adversarial strategies and evaluate the impact of the attacker's DER. Therefore, a trustworthy smart grid controller must monitor and manage each DERs operation for the entire smart-grid cyber-physical systems. 

To boost confidence in smart grid controllers and enable secure power transactions between DERs, it is crucial to develop a trustworthy artificial intelligence (AI) \cite{XAISurvey2023,TrustworthyAI1,TrustworthyAI2,TrustworthyAI3,TrustworthyAI4,munir2022neuro} mechanism for detecting potential cyber threats and measuring the severity of the risk. However, in order to establish trustworthy AI mechanisms for smart grid controllers, several technical metrics such as reliability, fairness, explainability, transparency, reproducibility, and accountability on threat detection must be needed to be fulfilled. 

In this work, we address the following research challenges:
\begin{itemize}
	\item How to proactively detect the potential cyber threat in a smart grid environment, where millions of DERs are connected and controlled by the smart grid controller?
	
	\item In order to guarantee a secure power transaction, how can the root cause of a possible cyber attack be identified in a smart grid controller? 
	
	\item When there are no apparent indicators to differentiate among types of attacks, how can the grid controller distinguish between the various attack types and measure the risk?
\end{itemize}
\begin{figure*}[!t]
	{
		\centering
		\includegraphics[width=0.95\textwidth]{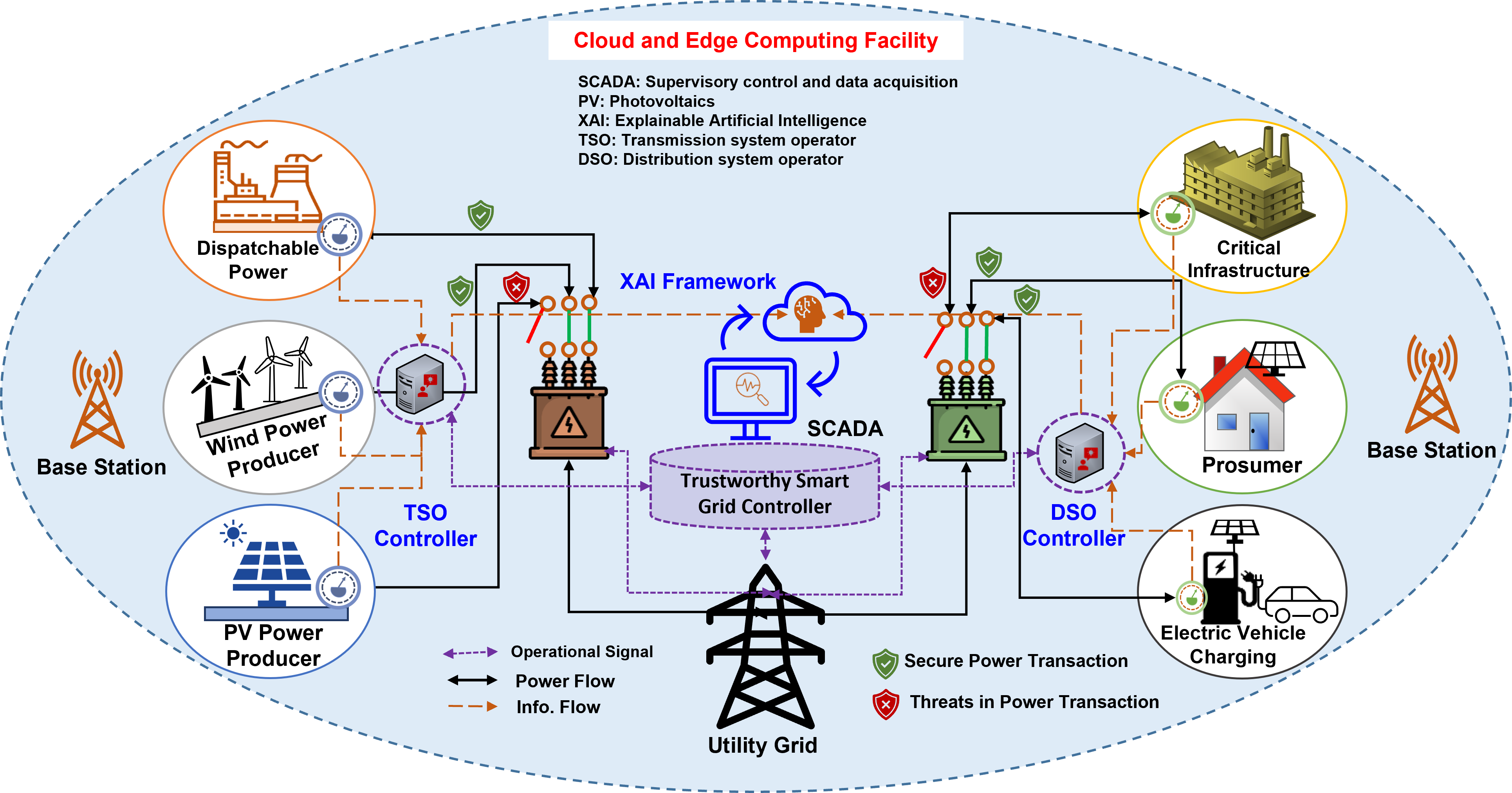}
		\caption{A system model of trustworthy smart grid controller.\label{fig: System_Model}}
	}
\end{figure*}
To address the above research challenges, in this work, a trustworthy artificial intelligence (AI) technique is studied for the purpose of proactively identifying and explaining the cyber risk brought on by the control/status message of DERs. We summarize our key contributions as follows:  
\begin{itemize}
	\item First, we design a system model of a trustworthy smart grid controller for assuring a secure power transmission among the distributed energy resources. Then, we formulate a decision problem that can detect the potential cyber threat in an adaptive smart grid environment while analyzing the root cause of such threat by determining the contribution among the features of a status message. 
	
	\item Second, we propose a trustworthy artificial intelligence framework to solve the formulated decision problem in smart grid controllers. In particular, we design an AI pipeline that is capable of facilitating the deployment of any AI algorithms for potential cyber threat detection, can analyze root causes based on Shapley \cite{Shap0,MunirXAIICC,Shap1,Shap2XAI} value interpretation, and dynamically quantifying the risk of an attack based on Ward's minimum variance \cite{ward1963hierarchical} formula.
	
	\item Third, we implement the proposed framework in a simulation environment and tested it with the state-of-the-art cyber-physical system SCADA WUSTL-IIOT-2018 \cite{Dataset2018}.
	
	\item Finally, we do the analysis of the implemented framework in terms of reliability, transparency, fairness, accountability, and explainability of the proposed trustworthy artificial intelligence framework. We have found that the proposed framework can detect potential attacks with at least $99\%$ reliability while ensuring transparency, fairness, and explainability by analyzing the contribution of the feature on decisions and quantifying the risk of each potential attack.
\end{itemize}

The rest of the paper is organized as follows. Section \ref{sec:related} discusses some of the interesting related work. The system model and problem formulation of the considered trustworthy smart grid controller are described in Section \ref{sec:system}. The proposed trustworthy AI framework is presented in Section \ref{sec:TXAI} and experimental analysis is discussed in Section \ref{sec:experiment}. Finally, we conclude our discussion in Section \ref{sec:con}. A summary of notations is presented in Table \ref{notation}.

\section{RELATED WORK}
\label{sec:related}
The problem of cyber attack detection for smart grid infrastructure has been studied in \cite{SmartgridAttack1,SmartgridAttack2,SmartgridAttack5,MunirGridShepherd,SmartgridAttack3}. In \cite{SmartgridAttack1}, the authors proposed an unsupervised learning-based anomaly detection by measuring the statistical correlation among features. The work in \cite{SmartgridAttack2} studied a partially observable Markov decision process for cyber-attack detection and proposed a model free reinforcement learning (RL) to solve it. The authors in \cite{SmartgridAttack5}  investigated the problem of false data injection (FDI) attack detection for the smart grid by deploying a support vector machine (SVM), K nearest neighbor (KNN), and artificial neural network (ANN)-based supervised learning mechanism. Further, the authors in \cite{MunirGridShepherd} proposed a data-informed policy-based model-free RL scheme for detecting cyber threats and controlling DERs for connected or disconnected from the main grid based on the attack situation. Recently, the authors \cite{SmartgridAttack3} proposed a hybrid deep learning model by combining particle swarm optimization (PSO) and convolutional neural networks-long short-term memory (CNN-LSTM) for FDI detection in the smart grid environment.

However, these works \cite{SmartgridAttack1,SmartgridAttack2,SmartgridAttack5,MunirGridShepherd,SmartgridAttack3} do not investigate the problem of root cause analysis of a particular cyber-attack nor do they account the quantify the potential risk, when there are no apparent indicators to differentiate among types of attacks. Dealing with explainability, reliability, and fairness in the detection of cyber attacks and measuring cyber risk with fairness, transparency, and accountability is challenging due to the intrinsic nature of millions of distinct DERs in smart grid cyber-physical systems. Therefore, in this work, we propose a trustworthy AI framework that can proactively recognize and explain the cyber risk posed by the control/status message of DERs. 

\section{SYSTEM MODEL AND PROBLEM FORMULATION OF TRUSTWORTHY SMART GRID OPERATION}
\label{sec:system}
\subsection{System Model of Trustworthy Smart Grid Controller}
We consider a smart grid environment, where a set $\mathcal{D} = \left\{{1,2,\dots,d, \dots, D}\right\}$ of distinct distributed energy resources (DERs) \cite{SmartgridAttack4,munirMetaRL,munirRiskAL} such as dispatchable power sources, wind sources, photovoltaics sources, prosumer, critical infrastructure consumers, electric vehicle (EV) charging station, and so on are deployed (as seen in Figure \ref{fig: System_Model}). The transmission system operator (TSO) controller transfers energy from source DERs to the distribution system operator (DSO). The DSO controller is responsible to distribute energy among the consumer DERs such as critical infrastructure, EV charging, prosumers, and others. Considering a \emph{trustworthy smart grid controller (TSGC)} that can monitor and control the activities of each DER for the entire smart grid cyber-physical systems. We assume that, physically, the TSGC is an entity of the supervisory control and data acquisition (SCADA) system of the considered smart grid infrastructure.  

\begin{table*}[t!]
	\caption{Summary of Notations}
	\begin{center}
		\begin{tabular}{|c|c|}
			\hline
			\textbf{Notation} & \textbf{Description} \\
			\hline
			$\mathcal{D}$ & Set of distributed energy resources (DERs) \\
			\hline
			$d \in \mathcal{D}$ & Each control/status messages \\
			\hline
			$\mathcal{X} = \left\{ 0, 1,2,\dots,x, \dots, X \right\}$ & Set of attack types including trusted message  \\
			\hline
			$M$ & No. of features \\
			\hline
			$y_d \in \mathbf{Y}$ & Threat detection decisions variable \\
			\hline
			$\Upsilon_{m} \in \boldsymbol{\Upsilon}$ & Shapley coefficient (root cause) \\
			\hline
			$g(.)$ & Risk of severity  \\
			\hline
			$z_d$ & Tuple of features \\
			\hline
			$\omega_M$ & Weight of the model for $M$ features \\
			\hline 
			$Z^t$ &  Total number of control/status messages at time slot $t \in T$ \\
			\hline
		\end{tabular}
		\label{notation}
	\end{center}
\end{table*} 
In the considered system model, three kinds of communication and energy flows occur power transmission, information flow, and operational message flow. In Figure \ref{fig: System_Model}, the yellow dashed line indicates information flow, the black solid line represents energy flow, and the purple dashed line presents operational signals among the DERs.

In general \cite{Dataset2018,MunirGridShepherd}, the at first control/status messages are exchanged between the generator DERs and TSO controller before transacting energy to the DSO. Similarly, control/status messages are exchanged between DSO controllers and consumer DERs before distributing the energy to the consumer DERs and vice versa. However, it is essential to assure trust in DERs before establishing energy flow or transaction while control messages may carry threats in the considered smart grid. In this system model, we consider five types of potential cyber threats: 1) port scanner, 2) address scan, 3) device identification, 4) aggressive mode, and 5) exploit \cite{Dataset2018,MunirGridShepherd} in a DER control/status message $d \in \mathcal{D}$. Therefore, we define a set $\mathcal{X} = \left\{ 0, 1,2,\dots,x, \dots, X \right\}$ that includes $X$ cyber threats and $x=0$ represents trusted control/status message. Thus, $X$ cyber threats can occur in smart grid cyber-physical systems and are initiated by the DERs $\forall d \in \mathcal{D}$. Thus, mathematically, a binary indicator can represent the observed cyber threat of each DER control/status message $d \in \mathcal{D}$,      
\begin{equation} \label{eq:wsc_binary_dec}
	\begin{split}
		y_d=
		\begin{cases}
			1, & \text{if}\ x \in \mathcal{X}, x \neq 0, \\
			0, & \text{trustworthy},
		\end{cases}
	\end{split}
\end{equation} 

where $y_d=1$ denotes the control message of DER $d \in \mathcal{D}$ is an attack, and $0$ trustworthy.


The considered TSGC can observe the characteristics and behavior of each DER $d \in \mathcal{D}$ by analyzing each of the control/status messages. In which, each control/status message consists of the following features: source port $a$, the total number of packets $b$, the total number of bytes $c$, the number of packets at source DER $\alpha$, the number of packets sent to destination $\beta$, and the number of source byte size $\gamma$ \cite{Dataset2018,MunirGridShepherd}. Therefore, we represent these features as a tuple $z_d \colon (a, b, c, \alpha, \beta, \gamma)$, $d \in \mathcal{D}$. However, in a dynamic case, we consider $M$ features, then each tuple represents as $z_d \colon (z_{d1}, \dots, z_{dm}, z_{dM}))$. Then, for each time slot $t \in T$, the total number of control/status messages are presented as $Z^t$, where $\forall z_{dm} \in M \times Z^t$ and $T$ in a finite time domain.

In this system model, our goal is to proactively detect the potential cyber threat $y$ by analyzing each control/status message before executing the power transaction command. However, only detecting the potential cyber threat does not buy the system as a trustworthy smart grid control. Therefore, the system model must contain such a mechanism that can explain the root cause of the detected threat along with a confidence score. We consider a linear model that can predict the potential cyber threat $y$ of DER control message $d \in \mathcal{D}$. For each feature $m$, the model is defined as follows \cite{MunirXAIICC}:
\begin{equation} \label{eq:wsc_linear_model}
\begin{split}
	y_d=\hat{h}(z_{dm}) = \omega_0 + \omega_1 z_{d1} + \dots + \omega_M z_{dM},
\end{split}
\end{equation} 
where $\omega_M$ denotes a weight of the model (\ref{eq:wsc_linear_model}). Therefore, $z_{dm}$ presents a feature value, where $m=1, \dots, M$ and $\omega_m$ represents a weight of feature $m$. 

In this work, our aim is to find the reasoning behind a potential cyber attack from the control/status message of each DER. Therefore, we can define a contribution function of feature $m$ on $\hat{h}(z_{dm})$ \cite{Shap0,MunirXAIICC,Shap1,Shap2XAI}. Then, thus, we can formulate a score function based on the contribution of the attack decision and define it as follows \cite{Shap0,Shap1}: 
\begin{equation} \label{eq:wsc_feature_cont}
\begin{split}
	\Upsilon_{m}(\hat{h}(z_{dm})) = \omega_m z_{dm} - \mathbb{E}[\omega_m Z^t],
\end{split}
\end{equation} 
where $\mathbb{E}[\omega_m Z^t]$ represents an expectation of effect for feature $m$. 
Here, our goal is to detect potential cyber threat $\mathbf{Y} \in \forall y$ that is initiated by DER $d \in \mathcal{D}$ and find the root cause behind that decision $ \forall \Upsilon_{m} (\hat{h}(z_{dm})) \in \boldsymbol{\Upsilon}, m \in M $.  Therefore, we need to formulate a problem for a trustworthy smart grid controller that can not only detect the potential attack but also can find the root cause to build trust in such a decision.

\subsection{Problem Formulation of Trustworthy Smart Grid Controller}
In this section, we design a decision problem for the proposed trustworthy smart grid controller that can coordinate between TSO and DSO. In this formulation, we consider two decision variables: 1) threat detection decisions $y_d \in \mathbf{Y}$, and 2) quantifying root cause $\Upsilon_{m} \in \boldsymbol{\Upsilon}$ of such decisions that are affected by the $M$ features of the control/status message of each DER $d \in \mathcal{D}$. The objective is to minimize the square error between actual occurrence $\hat{y_d}$ and predicted occurrence $y_d$ of each control/status message $d \in \mathcal{D}$ in the smart grid controller. Thus, we formulate the decision problem of the smart grid controller as follows:   
\begin{subequations}\label{Opt_1_1}
\begin{align}
	\underset{y \in \mathbf{Y},\Upsilon_{m} \in \boldsymbol{\Upsilon}}\min \;
	&\;  \frac{1}{|\mathcal{D}| |T| }\sum_{t=1}^{T} \sum_{d=1}^{|\mathcal{D}|} \big(\hat{y_d} -   y_d\big)^2  \tag{\ref{Opt_1_1}}, \\
	\text{s.t.} \quad & \label{Opt_1_1:const1} \omega_m z_{dm} \le \mathbb{E}[\omega_m Z^t], \forall m \in (1, \dots, M),\\
	&\label{Opt_1_1:const2} \hat{y_d} \ge \omega_0 + \omega_1 z_{d1} + \dots + \omega_M z_{dM}, m \in M, d \in \mathcal{D}, \\
	&\label{Opt_1_1:const3} \Upsilon_{m} \le \sum_{m=1}^{M}  (\omega_m z_{dm} - \mathbb{E}[\omega_m Z^t] ) ,\forall \Upsilon_{m} \in \boldsymbol{\Upsilon}, \\
	&\label{Opt_1_1:const4} y_d|\mathcal{D}| \le |\mathbf{Y}|, \\
	& \label{Opt_1_1:const5} y_d \in \left\lbrace0,1 \right\rbrace, \forall d \in \mathcal{D}.
\end{align}
\end{subequations}
Constraint (\ref{Opt_1_1:const1}) ensures that the contribution of feature $m \in \mathcal{M}$ must be smaller or equal to the expectation of effect $m$ for a control/status message of DER $d \in \mathcal{D}$. Constraint (\ref{Opt_1_1:const2}) ensures the linear predicted model (\ref{eq:wsc_linear_model}) never exceeds the actual value during estimation. We establish a coupling between prediction and contribution (i.e., effect) in constraint (\ref{Opt_1_1:const3}). Finally, constraints (\ref{Opt_1_1:const4}) and (\ref{Opt_1_1:const5}) assure that the total number of threats prediction does not bigger than the number of status messages and each status message $d \in \mathcal{D}$ only has one decision, respectively. 

The formulated problem (\ref{Opt_1_1}) leads to a combinatorial optimization problem in both the time and space domain due to potential threats are depended on time and are characterized by the nature of each DER status message. The problem (\ref{Opt_1_1}) is hard to solve in polynomial time; however, it can be solved through heuristic approximation. In this work, we propose a Shapley-based \cite{Shap0,MunirXAIICC,Shap1,Shap2XAI} trustworthy artificial intelligence framework to solve the formulated problem. The proposed framework can proactively detect cyber threats that are generated by DER, find the root cause by analyzing the features of a status message, and elaborate on the severity of the cyber risk.

\section{PROPOSED TRUSTWORTHY ARTIFICIAL INTELLIGENCE FRAMEWORK}
\begin{figure*}[!t]
{
\centering
\includegraphics[width=0.95\textwidth]{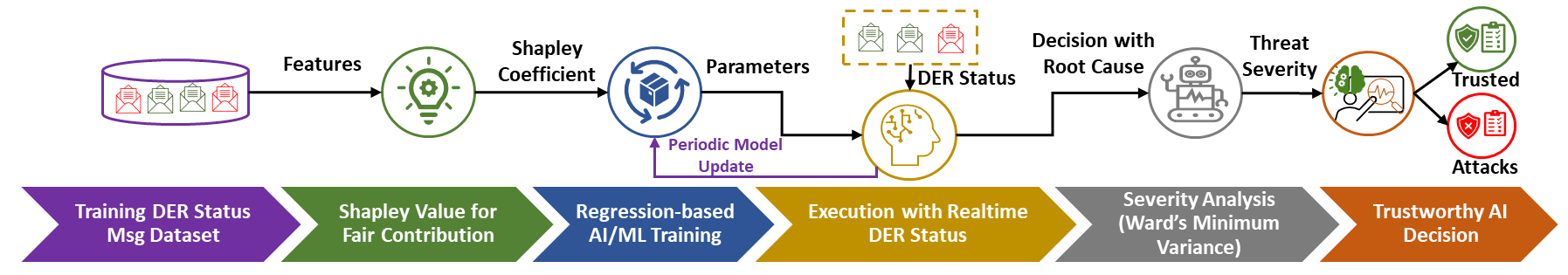}
\caption{The proposed trustworthy AI framework for the smart grid controller.\label{sol}}
}
\end{figure*}
\label{sec:TXAI}
We illustrate the high-level system design of the proposed trustworthy AI framework in Figure \ref{sol}. In particulate, Figure \ref{sol} demonstrated the overall AI pipeline for enabling a trustworthy smart grid controller. The developed framework incorporates a regression-based predictive mechanism for cyber threat detection, a Shapley-based explainer for root cause analysis, and Ward's minimum variance-based hierarchical clustering for characterizing threat severity. 

The root cause of a status message $d$ in feature $m$ can be estimated as follows:
\begin{equation} \label{eq:wsc_feature_contribution}
\begin{split}
\sum_{m=1}^{M}\Upsilon_{m}(\hat{h}(z_{dm})) = \sum_{m=1}^{M}  (\omega_m z_{dm} - \mathbb{E}[\omega_m Z^t] )\\ = \hat{h}(z_{dm}) - \mathbb{E}[\hat{h}(Z^t)],
\end{split}
\end{equation}
where $\omega_m$ is the weight of contributing feature $m$ and $z_{dm}$ denote the each feature $m$ in a feature tuple $z$. $\mathbb{E}[\omega_m Z^t]$ represents expected effect by the feature $m$ in DER status message $d$. Therefore, the root cause of a particular threat detection decision can be calculated as follows:  
\begin{equation} \label{eq:wsc_shap_cal}
\begin{split}
\Upsilon_{m} = \;\;\;\;\;\; \;\;\;\;\;\; \;\;\;\;\;\; \;\;\;\;\;\; \;\;\;\;\;\; \;\;\;\;\;\; \;\;\;\;\;\; \;\;\;\;\;\; \;\;\;\;\;\; \;\;\;\;\;\; \;\;\;\;\;\; \;\;\;\;\;\; \;\;\;\;\;\; \\ \frac{1}{|\mathcal{M}|} \sum_{\mathcal{M} \subseteq \mathcal{D} \setminus \left\{ m \right\}} \big[|\mathcal{M}| \times\begin{array}{c} |\mathcal{D}| \\|\mathcal{M}| \end{array} \big]^{-1} [\hat{h}(z_{dm})_{\mathcal{M}}  - \hat{h}(z_{dm})_{\mathcal{M} \setminus m} ] \\ = \frac{marginal \; contribution \; of \; m}{number \; of \; coalitions \; \mathcal{M} \subseteq \mathcal{D} \setminus \left\{ m \right\} }, 
\end{split}
\end{equation}
where $y_d=\hat{h}(z_{dm})$ in (\ref{eq:wsc_linear_model}). Thus, $\Upsilon_{m}$ provides the root cause of the potential cyber attack $y_d=\hat{h}(z_{dm})$ for each feature $m \in \mathcal{M}$ of DER status message $d \in \mathcal{D}$.

\begin{algorithm}[!t]
\caption{An algorithm for trustworthy smart grid controller}
\label{alg:TSGC}
	\begin{algorithmic}[1]
	\renewcommand{\algorithmicrequire}{\textbf{Input:}}
	\renewcommand{\algorithmicensure}{\textbf{Output:}}
	\REQUIRE $ \forall m \in \mathcal{M}, \forall d \in \mathcal{D}, Z^t $, $z_d \colon (a, b, c, \alpha, \beta, \gamma)$, $d \in \mathcal{D}$, $\forall \hat{y} \in \hat{Y}$
	\ENSURE  $y_d \in \mathbf{Y}$, $\Upsilon_{m} \in \boldsymbol{\Upsilon}$, $g(.)$
	\WHILE{$T \neq T_{max}$}
	\FOR {$\forall d \in \mathcal{D}$} 
	\STATE Select a feature $z_{dm} \in Z^t$
	\FOR {$\forall m \in \mathcal{M}$}
	\STATE Calculate: $\omega_m z_{dm}$ and $\mathbb{E}[\omega_m Z^t]$
	\IF{$\omega_m z_{dm} \le \mathbb{E}[\omega_m Z^t]$ (Constraint (\ref{Opt_1_1:const1}))}
	\STATE Calculate: $(\hat{h}(z_{dm}) - \mathbb{E}[\hat{h}(Z^t)] )$
	\STATE Estimate: $\omega_0 + \omega_1 z_{d1} + \dots + \omega_M z_{dM}$
	\STATE Calculate: $\Upsilon_{m} \gets \frac{1}{|\mathcal{M}|} \sum_{\mathcal{M} \subseteq \mathcal{D} \setminus \left\{ m \right\}} \big[|\mathcal{M}| \times\begin{array}{c} |\mathcal{D}| \\|\mathcal{M}| \end{array} \big]^{-1} [\hat{h}(z_{dm})_{\mathcal{M}}  - \hat{h}(z_{dm})_{\mathcal{M} \setminus m} ] $ \COMMENT{Using (\ref{eq:wsc_shap_cal})}
	\ENDIF
	\ENDFOR
	\IF{(\ref{Opt_1_1:const2}) and (\ref{Opt_1_1:const3})}
	\STATE Evaluate  $\frac{1}{|\mathcal{D}| |T| }\sum_{t=1}^{T} \sum_{d=1}^{|\mathcal{D}|} \big(\hat{y_d} -   y_d\big)^2 $ \COMMENT{Generic loss function evaluation for AI models} 
	\STATE Get Threat Decision: $y_d$ 
	\STATE Get Explanation: $\Upsilon_{m}$
	\ENDIF
	\ENDFOR
	\STATE $\mathbf{Y} \gets y$, $\boldsymbol{\Upsilon} \gets \Upsilon_{m}$
	\STATE Severity Analysis: $g (\boldsymbol{\Upsilon}, \mathcal{X})$ \COMMENT{Attack risk analysis using (\ref{eq:ward_expand})}
	\ENDWHILE
	\STATE Return: $\mathbf{Y}, \boldsymbol{\Upsilon}$, $g (\boldsymbol{\Upsilon}, \mathcal{X})$  
	\end{algorithmic}
\end{algorithm}

Now, we need to characterize the severity of the potential threat risk. However, the challenge here is to adapt to the unknown behavior of a potential cyber threat in smart grid controllers. We consider each $X_i$ can represent a disjoint type of cyber threat or trusted energy transaction control/status message, where $\forall X_i \in \mathcal{X}$. Therefore, for two instance (i.e., $i$ and $i+1$), we can define Ward's minimum variance \cite{ward1963hierarchical} formula as follows:
\begin{equation} \label{eq:ward_main}
\begin{split}
g (\boldsymbol{\Upsilon}, \mathcal{X}) = \frac{|X_i| |X_{i+1}|}{|X_i| \cup |X_{i+1}|} ||\mu_{X_i} - \mu_{X_{i+1}}||^{2}_{\sim \forall  \Upsilon_{m} \in \boldsymbol{\Upsilon}}, 		
\end{split}
\end{equation}
where $\mu_{X_i}$ and $\mu_{X_{i+1}}$ denote centroid of $|X_i|$ and $X_{i+1}$. We can rewrite (\ref{eq:ward_main}) as follows:
\begin{equation} \label{eq:ward_expand}
\begin{split}
g (\boldsymbol{\Upsilon}, \mathcal{X}) = \sum_{x \in {X_i} \cup {X_{i+1}}}|| x - \mu_{{X_i} \cup {X_{i+1}}}  ||^{2}_{\sim \forall  \Upsilon_{m} \in \boldsymbol{\Upsilon}} - \\ \sum_{x \in {X_i} }|| x - \mu_{{X_i}}  ||^{2}_{\sim \forall  \Upsilon_{m} \in \boldsymbol{\Upsilon}} - \sum_{x \in {X_{i+1}} }|| x - \mu_{{X_{i+1}}}  ||^{2}_{\sim \forall  \Upsilon_{m} \in \boldsymbol{\Upsilon}}.	
\end{split}
\end{equation}

An algorithmic procedure of the proposed mechanism for the trustworthy smart grid controller is shown in Algorithm \ref{alg:TSGC}. Physically, Algorithm \ref{alg:TSGC} will be executed by the smart grid controller. The input of the Algorithm \ref{alg:TSGC} will be control/status messages that are sent by DERs $\forall d \in \mathcal{D}$ and each message contains $|\mathcal{M}|$ features. The Algorithm \ref{alg:TSGC} can be run in a finite time domain and is also capable of working as a watchdog in an infinite time domain. All of the received control/status messages will be filtered as a potential cyber threat or trustworthy $y_d$ and quantifying the root cause of that decision $\boldsymbol{\Upsilon}$. Based on that findings, the Algorithm \ref{alg:TSGC} characterizes the risk of severity $g(.)$ among the decisions. Lines from $4$ to $11$ estimate the predicted decision based on the regression and calculate each feature's contribution $\Upsilon_{m} \in \boldsymbol{\Upsilon}$ based on the Shapley value interpretation. In particular, line $8$ executes the regression model and line $9$ calculates the contribution of each feature in Algorithm \ref{alg:TSGC}. Line $13$ evaluates a loss function for regression-based AI model training such as random forest, extra tree, adaboots, and so on. Finally, cyber attack severity is analyzed in line $19$ of Algorithm \ref{alg:TSGC}. The computational complexity of the proposed Algorithm \ref{alg:TSGC} relies on the number of features $|\mathcal{M}|$ of a control/status message in a smart grid controller. The average case  Computational complexity of the proposed Algorithm \ref{alg:TSGC} leads to $\mathcal{O}(2^{|\mathcal{D}| \times |\mathcal{M}|} + |\mathcal{X}|^2 \log |\mathcal{X}| )$, where $|\mathcal{X}|$ is the number of disjoint types of cyber threats in smart grid controller.

\section{EXPERIMENTAL ANALYSIS}
\label{sec:experiment}
\begin{figure}[!t]
	\centering
	\includegraphics[scale=0.4]{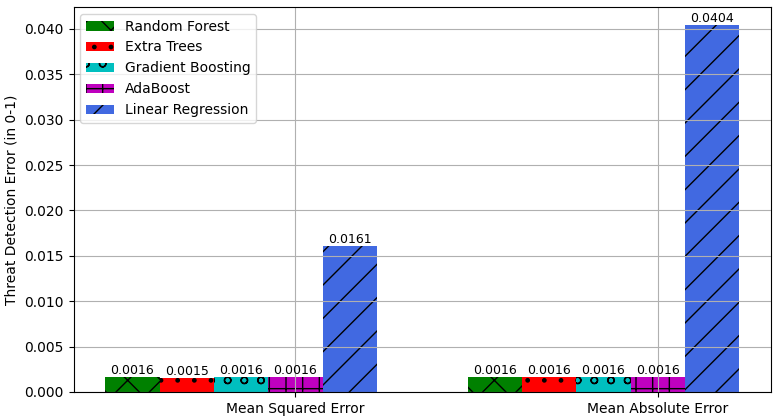}
	\caption{Trend analysis of mean squared error and mean absolute error among the ensemble-based regression AI/ML models during execution (i.e., testing).}
	\label{Error}
\end{figure}
\subsection{Experiment Setup}
\begin{table}[t!]
\caption{Summary of Experimental Setup}
\begin{center}
	\begin{tabular}{|c|c|}
				\hline
				\textbf{Description} & \textbf{Value} \\
				\hline
				No. of DERs status/control message sessions & $2000$  \cite{Dataset2018}\\
				\hline
				No. of training sessions & $1400$  \cite{Dataset2018}\\
				\hline
				No. of testing sessions & $600$ \cite{Dataset2018} \\
				\hline
				Types of attacks & $5$ \cite{Dataset2018} \\
				\hline
				No. of control/status message features & $6$  \cite{Dataset2018}\\
				\hline
				Cluster affinity & Euclidean \\
				\hline
				Cluster linkage & Complete \\
				\hline
				Cluster method & Ward \\
				\hline
			\end{tabular}
			\label{exp_steup}
		\end{center}
	\end{table} 
The proposed trustworthy AI framework for a smart grid controller is implemented in Python platform \cite{EnsemblePython,shapPython,ClusterPython}. In particular, we have developed several ensemble machine learning methods \cite{EnsemblePython} such as Extra Tree, Random Forest, Gradient Boosting, AdaBoost, and Linear Regression for testing the proposed trustworthy smart grid controller. We have utilized the SHAP \cite{shapPython} Python library for calculating Shapley value-based root cause analysis of the potential cyber attack. Then, we used Scikit-learn library to implement the Agglomerative \cite{ClusterPython} clustering for severity characterization of the potential cyber risk. A summary of the experimental setup is given in Table \ref{exp_steup}.

The implemented framework is tested using the state-of-the-art cyber-physical system SCADA WUSTL-IIOT-2018 \cite{Dataset2018}. In this dataset, each control/status message consists of $6$ features along with the true label. The features include the source port, the total number of packets, the total number of bytes, the number of packets at source DER, the number of packets sent to the destination, and the number of source bytes. We have considered $2000$ DER control/status messages and divided these into $70\%$ for training and the rest of them are used for testing. In this dataset, five types of potential cyber threats are considered: 1) port scanner, 2) address scan, 3) device identification, 4) aggressive mode, and 5) exploit; these are labeled into one category. Therefore, there are no given clues that can differentiate between the types of attacks. To overcome such challenges, we have utilized the concept of unsupervised learning for measuring the severity of cyber risk.

\begin{figure*}[t!] 
	\centering
	\begin{subfigure}[t]{0.45\textwidth}
		\centering
		\includegraphics[height=2.0in]{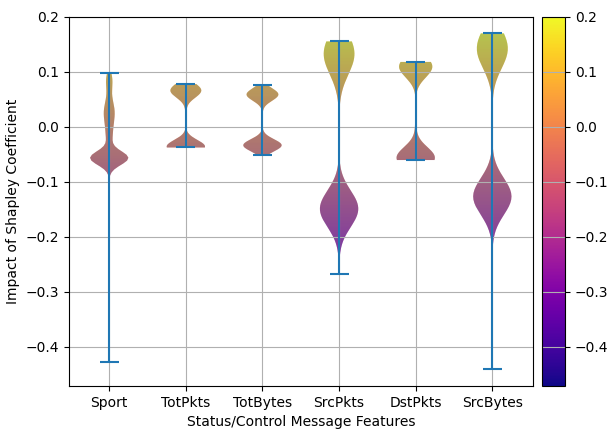}
		\caption{Contribution of each feature on attack detection from DER control/status messages.}
		\label{Shapley_feature_abnormal}
	\end{subfigure}
	~
	\begin{subfigure}[t]{0.45\textwidth}
		\centering
		\includegraphics[height=2.0in]{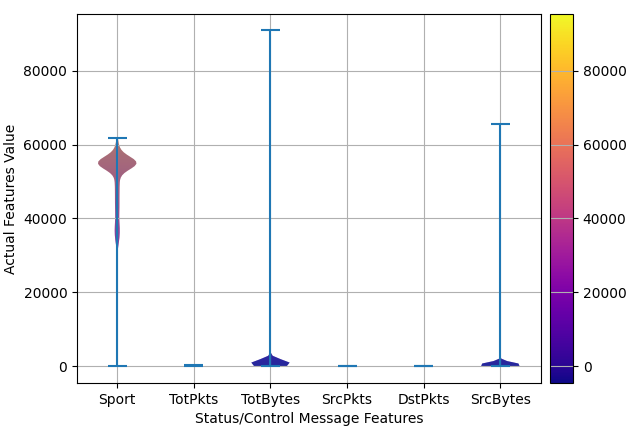}
		\caption{Features correlation of DER messages in WUSTL-IIOT-2018 dataset \cite{Dataset2018}.}
		\label{actual_feature_value}
	\end{subfigure}%
	\caption{Explanation of the impact of Shapley value coefficient in cyber attack detection.}
\end{figure*}

\begin{figure*}[t!]
	\centering
	\begin{subfigure}[t]{0.45\textwidth}
		\centering
		\includegraphics[height=2.0in]{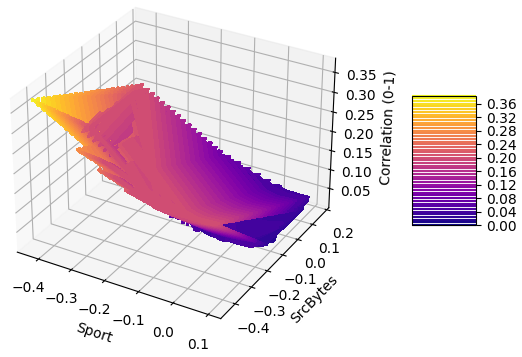}
		\caption{Effect of generated bytes in source port of DERs on attack detection.}
		\label{sport_correlation}
	\end{subfigure}%
	~ 
	\begin{subfigure}[t]{0.45\textwidth}
		\centering
		\includegraphics[height=2.0in]{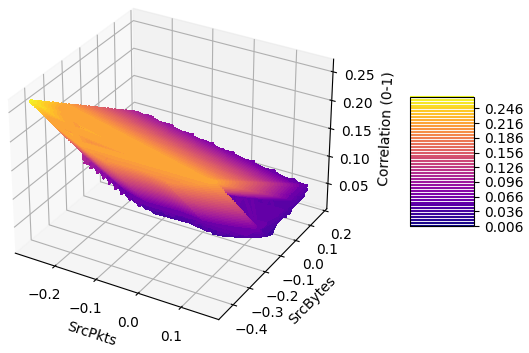}
		\caption{Effect of generated bytes and source packet of DERs on attack detection.}
		\label{srcPkts_correlation}
	\end{subfigure}
	\caption{Characterizing evidence on most prominent contributed features in DER control/status messages.}
\end{figure*}
	
\subsection{Results and Discussion}
In this experiment, we focus on establishing a trustworthy smart grid controller by validating the proposed trustworthy AI framework for SCADA. Thus, in order to establish a trustworthy AI model, we need to fulfill a few of the metrics \cite{TrustworthyAI1,TrustworthyAI2,TrustworthyAI3,TrustworthyAI4,munir2022neuro} such as \emph{reliability}, \emph{transparency}, \emph{fairness}, \emph{accountability}, \emph{reproducibility}, and \emph{explainability}. Therefore, we justify the proposed trustworthy AI framework in the following subsections.  
	\subsubsection{Reliability}
	\begin{table*}[t!]
		\caption{Reliability Analysis on AI Decisions (Score between $0$ to $1$)}
		\begin{center}
			\begin{tabular}{|c|c|c|c|c|c|c|}
						\hline
						\textbf{AI Methods} & \textbf{TN} & \textbf{FP} & \textbf{FN} & \textbf{TP} & \textbf{$R^2$ Training Score} & \textbf{$R^2$ Test Score} \\
						\hline
						Random Forest & $0.996$ & $0.003$ & $0.0$ & $1.0$ & $0.997$ & $0.993$  \\
						\hline
						Extra Trees & $0.996$ & $0.003$ & $0.0$ & $1.0$ & $0.998$ & $0.993$   \\
						\hline
						Gradient Boosting & $0.996$ & $0.003$ & $0.0$ & $1.0$ & $0.997$ & $0.993$   \\
						\hline
						AdaBoost & $0.996$ & $0.003$ & $0.0$ & $1.0$ & $0.997$ & $0.993$  \\
						\hline
						Linear Regression & $0.996$ & $0.003$ & $0.0$ & $1.0$ & $0.946$ & $0.935$   \\
						\hline
					\end{tabular}
					\label{reliability}
				\end{center}
			\end{table*}

In order to measure the reliability of the proposed trustworthy AI framework, we consider a set of metrics \cite{Metrics} such as true negative (TN), false positive (FP), false negative (FN), true positive (TP), $R^2$ regression score for training and testing, mean squared error (MSE) and mean absolute error (MAE). In Table \ref{reliability}, we have presented TN, FP, FN, and TP rates for $600$ DER control/status messages, where we have found $0.003$ and $1.0$ as the FP and TP rates, respectively for all of the AI/ML methods. Table \ref{reliability} also evidenced a higher $R^2$ score (i.e., coefficient of determination) for all cases. Further, Figure \ref{Error} demonstrates the reliability of the defined objective function \ref{Opt_1_1} (i.e., mean squared error) during testing, where we achieved a minimum error rate $0.0015$ by the Extra Tree model in the trustworthy AI framework.

\subsubsection{Transparency, Fairness, and Accountability}
To assure transparency, fairness, and accountability of the proposed trustworthy AI framework, we have demonstrated the impact of the Shapley value coefficient for DER cyber attack detection in Figures  \ref{Shapley_feature_abnormal}. In particular, it is clearly understandable in Figure \ref{Shapley_feature_abnormal} that the source port (Sport), the size of the source packet (SrcPkts), and the number of source bytes (SrcBytes) have more contribution for detecting trusted and threat DER messages. However, feature correlation (in \ref{actual_feature_value}) of the raw data does not convey the right insight. As a result, the proposed Shapley value-based feature contribution in attack detection assures fairness and transparency in the AI model's decision. 
			
We have illustrated around $35\%$ effect of generated bytes (SrcBytes) in source port (Sport) in Figure \ref{sport_correlation} while Figure \ref{srcPkts_correlation} depicts that generated bytes (SrcBytes) and source packet (SrcPkts) have around $25\%$ effects on DERs' attack detection. This evidence assures the accountability of the proposed trustworthy AI framework.
			
\subsubsection{Reproducibility and Explainability}
\begin{figure*}[t!]
\centering
\begin{subfigure}[t]{0.42\textwidth}
		\centering
		\includegraphics[height=1.75in]{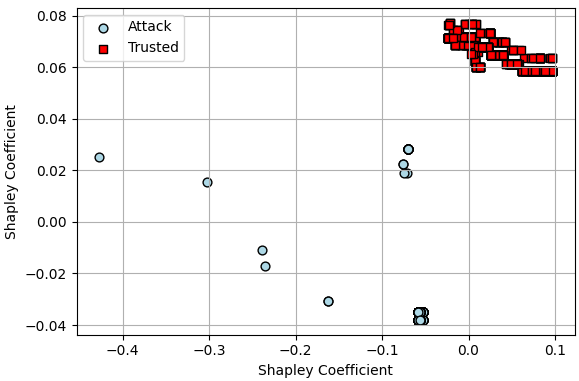}
		\caption{Shapley value-based explanation of threat and trusted DER messages.}
		\label{cluster_two}
	\end{subfigure}%
	~ ~ ~ ~ ~ 
	\begin{subfigure}[t]{0.42\textwidth}
		\centering
		\includegraphics[height=1.75in]{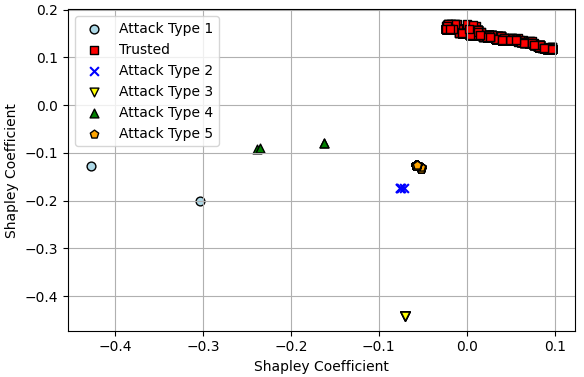}
		\caption{Characterization of the severity of five types of cyber attacks in unknown labels.}
		\label{cluster_six}
	\end{subfigure}
	\caption{Outcomes of agglomerative (hierarchical) clustering for the severity explanation.}
\end{figure*}

\begin{figure*}[t!]
	\centering
	\includegraphics[scale=0.5]{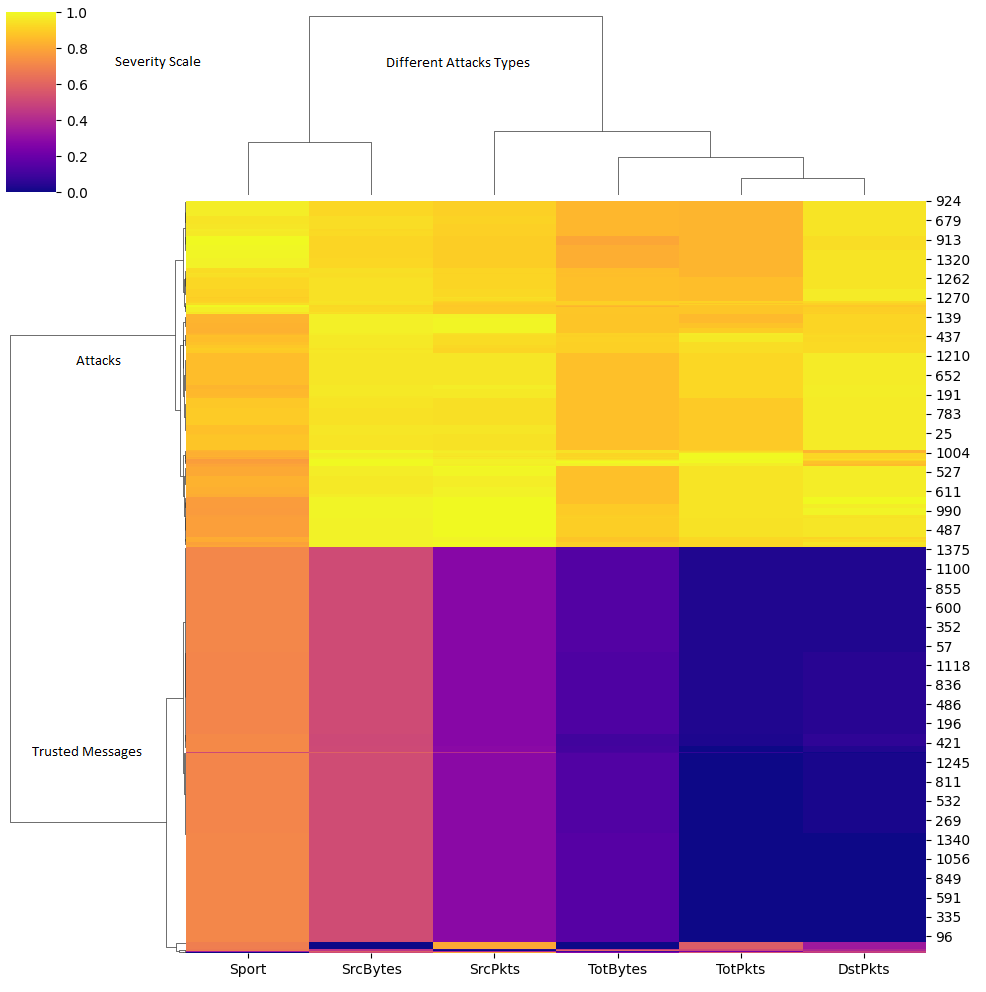}
	\caption{Risk quantification and explanation among the cyber threats in smart grid controller by Ward's minimum variance-based hierarchical clustering.}
	\label{wards_min}
\end{figure*}

We assure the reproducibility and explainability of the proposed trustworthy AI framework by incorporating Ward's minimum variance-based hierarchical clustering scheme. In particular, the trustworthy AI framework explains the Shapley coefficient-based root cause and severity of the potential cyber risk that are shown in Figures \ref{cluster_two}, \ref{cluster_six}, and \ref{wards_min}. The explanation between trusted DER messages and potential attacks is shown in Figures \ref{cluster_two}, where we can clearly observe that the potential threats do not follow any particular characteristics. Therefore, we differentiate among the unknown types of cyber threats in Figure \ref{cluster_six} that can interpret the root cause of such threats based on the Shapley coefficient.
			
The trustworthy smart grid controller can observe the severity of potential cyber risk from the DER status/control messages. In Figure \ref{wards_min}, we quantify the severity of cyber risk during the execution. In Figure \ref{wards_min}, the x-axis represents the Shapley coefficient of features and the y-axis is control messages. Figure \ref{wards_min} indicates that the variation of Sport and SrcByte introduced a high risk of potential attack. To this end, the experimental analysis establishes the proposed AI framework as a trustworthy AI mechanism that can satisfy reliability, transparency, fairness, accountability, and explainability.

\section{CONCLUSION}
\label{sec:con}
In this paper, we have introduced a new trustworthy artificial intelligence framework for assuring trusted power transmission among the DERs. The proposed trustworthy AI framework can proactively predict and explain the root cause of potential cyber-attacks in smart grid controllers while measuring the severity of such attacks. The developed trustworthy AI framework can select prominent features by deploying Shapley value interpretation and train numerous regression-based ML/AI techniques by executing the mean squared error loss function. Further, it can interpret the root cause and severity of the potential cyber-attacks by employing Ward's minimum variance-based hierarchical clustering. Experimental analysis shows the efficacy of the proposed framework in terms of reliability, fairness, and explainability. That ensures the trustworthiness of the proposed AI framework. In the future, we will investigate a neuro-symbolic AI scheme for autonomous recovering system faults in a smart grid framework.

	
\bibliographystyle{IEEEtran}
\bibliography{demobib}

\section*{AUTHOR BIOGRAPHIES}
\noindent {\bf MD. SHIRAJUM MUNIR} received the Ph.D. degree in computer engineering from Kyung Hee University (KHU), South Korea, in 2021. He is currently working as a Post-doctoral Research Associate at Virginia Modeling, Analysis, and Simulation Center, Old Dominion University, USA. His research interests include machine learning, data science, trustworthy artificial intelligence and stochastic models for wireless network resource management, sustainable edge computing, healthcare, industrial IoT, intelligent IoT network management, future internet, and resilient smart grid.\\

\noindent {\bf SACHIN SHETTY} received the Ph.D. degree in modeling and simulation from Old Dominion University, in 2007. He is currently the Associate Director of the Virginia Modeling, Analysis, and Simulation Center, Old Dominion University. He holds a joint appointment as an Professor with the Department of Computational, Modeling, and Simulation Engineering. His research interests include the intersection of computer networking, network security, and machine learning. \\

\noindent {\bf DANDA B. RAWAT} is the Executive Director, Research Institute for Tactical Autonomy (RITA) - a University Affiliated Research Center (UARC) of the US Department of Defense, Associate Dean for Research and Graduate Studies, a Full Professor in the Department of Electrical Engineering Computer Science (EECS), Founding Director of the Howard University Data Science Cybersecurity Center, Director of DoD Center of Excellence in Artificial Intelligence Machine Learning (CoE-AIML) at Howard University, Washington, DC, USA. \\

\end{document}